\documentclass[12pt,preprint]{emulateapj}

\slugcomment{To Appear in the Astrophysical Journal}

\shorttitle{Metallicity of Sub-DLAs}
\shortauthors{Kulkarni et al.}

\begin{document}

\title{The Role of Sub-Damped Lyman-alpha Absorbers in the Cosmic Evolution of 
Metals}

\author{V. P. Kulkarni\altaffilmark{1}, P. Khare\altaffilmark{2}, 
C. P\'eroux\altaffilmark{3}, 
D. G. York \altaffilmark{4,5}, 
J. T. Lauroesch \altaffilmark{6}, J. D. Meiring\altaffilmark{1}}

\altaffiltext{1}{Department of Physics and Astronomy, University of South Carolina, 
Columbia, SC 29208; E-mail: kulkarni@sc.edu}
\altaffiltext{2}{Department of Physics, Utkal University, Bhubaneswar, 751004, India}
\altaffiltext{3}{European Southern Observatory, Garching-bei-Munchen, Germany}
\altaffiltext{4}{Department of Astronomy and Astrophysics, University of Chicago, 
Chicago, IL 60637}
\altaffiltext{5}{Also, Enrico Fermi Institute}
\altaffiltext{6}{Department of Physics and Astronomy, 
University of Louisville, Louisville, KY 40292; Also, Visiting Scholar, Northwestern University.}

\begin{abstract}

Observations of low mean metallicity of damped Lyman-alpha (DLA) quasar 
absorbers at all redshifts studied appear to contradict the predictions 
for the global mean interstellar metallicity in galaxies from cosmic chemical 
evolution models. On the other hand, a number of 
metal-rich sub-DLA systems have been identified recently, and the fraction 
of metal-rich sub-DLAs appears to be considerably larger than that of metal-rich DLAs, 
especially at $z < 1.5$. In view of this, 
here we investigate the evolution of metallicity in sub-DLAs. 
We find that the mean Zn metallicity of the observed sub-DLAs may 
be higher than that of the observed DLAs, especially at low redshifts, 
reaching a near-solar level at $z \la 1$. This trend does not appear to be 
an artifact of sample selection, the use of Zn, the use 
of $N_{\rm H I}$-weighting, or observational sensitivity. While a bias against very low metallicity could be present in the sub-DLA sample in some situations, 
this cannot explain the difference between the DLA and 
sub-DLA metallicities at low $z$. The primary reason for the difference between the DLAs and sub-DLAs appears to be the dearth of metal-rich DLAs. We estimate the sub-DLA contribution to the 
total metal budget using measures of their metallicity and comoving gas 
density. These calculations suggest that at $z \la 1$, the contribution of 
sub-DLAs to the total metal budget may be several times that of DLAs.
At higher redshifts also, there are indications that the 
sub-DLAs may contribute significantly to the cosmic metal budget. 

\end{abstract}


\section{Introduction}

As the cycle of stellar birth and death progresses in galaxies, 
the interstellar gas 
is expected to become increasingly metal-rich. A fundamental prediction of 
most chemical evolution models is that the mass-weighted 
mean interstellar metallicity of galaxies   
should rise from a low value at $z > 2$ to a near-solar 
value at $z=0$  (e.g., Pei et al. 1999; 
Somerville et al. 2001; Cora et al. 2003). Indeed, the mass-weighted mean metallicity of local 
galaxies is close to solar (e.g., Kulkarni \& Fall 2002; 
Fukugita \& Peebles 2004). How was this present level of metallicity reached?
A direct way to study this enrichment history is through the 
absorption lines superposed by galaxies on the spectra of background quasars, 
in particular the DLAs (defined to have neutral hydrogen column densities 
log $N_{\rm HI} \ge 20.3$) and the sub-DLAs 
(defined to have $19. 0 \le$ log $N_{\rm HI} < 20.3$; P\'eroux et al. 2003). 
These absorbers 
contain sufficient neutral 
gas mass to account for a significant fraction of the visible stellar mass in 
present-day galaxies 
(e.g., Storrie-Lombardi \& Wolfe 2000; P\'eroux et al. 2003, 2005; 
Prochaska et al. 2005). Indeed, hydrodynamical simulations 
naturally produce DLAs from cool gas accumulating at  
centers of cold dark matter (CDM) halos and predict DLAs to be closely 
associated  with star forming regions (e.g., 
Haehnelt et al. 2000; Cen et al. 2003; Nagamine et al. 
2004).  
 
Previous studies of element abundances in high-$N_{\rm H I}$ quasar 
absorbers have focussed primarily on DLAs, especially those at high 
redshift (e.g., Prochaska et al. 2003b). Recently, we have started to expand 
the low-$z$ DLA Zn data and tripled the $z < 1.5$ sample so far 
(Khare et al. 2004; Kulkarni et al. 2005; Meiring et al. 
2006a, 2006b; P\'eroux et al. 2006b). These studies find that 
the DLA mean metallicity stays substantially subsolar even at low 
redshifts and increases relatively slowly  (see also Prochaska et al. 2003a). 
Furthermore, based on deep 
emission-line imaging studies (e.g. Kulkarni et al. 2006 and references 
therein), the star formation rates (SFRs) of a large fraction of DLAs appear to fall far below the 
predictions from the cosmic star formation 
history (inferred from galaxy imaging surveys such as the Hubble Deep Field, e.g., Madau et al. 1998). 
A missing metals problem has also been noted in high-$z$ DLAs from SFR 
estimates based on C II* absorption (Wolfe et al. 2003).  
These results could potentially present important
challenges to our understanding of galaxy evolution. 

Why do DLA data disagree with the mean 
metallicity in galaxies predicted from the global star formation 
history? 
Smoothed particle hydrodynamics (SPH) simulations 
(Nagamine et al. 2004) indicate that true DLA metallicities could be
1/3 solar at $z=2.5$ and higher at lower $z$. Has the DLA 
metal content been underestimated, or is there a larger quantity of metals 
in a hitherto unexplored absorber population? While dust selection 
or metallicity gradients can lead to a flat 
metallicity-redshift relation, the extent of these effects is still 
uncertain (e.g., Boiss\'e et al. 1998;  
Vladilo \& P\'eroux 2005; Ellison et al. 2005; Akerman et al. 2005; 
Chen et al. 2005; Zwaan et al. 2005). It has 
been suggested recently that dust obscuration is not significant and 
that metal-rich DLAs are inherently rare (Meiring et al. 2006a; 
Khare et al. 2006, hereafter K06; 
Herbert-Fort et al. 2006). In view of this, 
it is important to explore other possible sites for the missing metals. 

Over the past two years, we have been investigating the metallicities of 
sub-DLAs at $z < 1.5$ (Khare et al. 2004; P\'eroux et al. 2006a, 2006b; 
Meiring et al. 2006b). These studies have led to the discovery of five 
solar or supersolar metallicity sub-DLAs at $0.7 < z < 1.5$. Before these 
studies began, only a single  metal-rich sub-DLA was identified 
(Pettini et al 2000). Indeed, 
five of the nine sub-DLAs with Zn data at $z < 1$ and six of the 17 sub-DLAs at 
$z < 1.5$ have solar or supersolar metallicity. Despite the small number of 
measurements (caused by the focus on DLAs in the past), the large fraction of 
metal-rich sub-DLAs suggests that  log $N_{\rm H I} \approx 20.0$ may be 
the optimum range 
for revealing metal-rich systems, since such $N_{\rm H I}$ is high enough to 
have a significant metal column density and low enough to make dust attenuation 
less important (Lauroesch et al. 1996; Vladilo \& P\'eroux 2005; P\'eroux et al. 2005, 2006a). 
To investigate this further, here we examine the evolution of the 
observed amount of metals in neutral gas in sub-DLAs and their contribution 
to the cosmic metal budget. 

\section{Metallicity Evolution of DLAs and Sub-DLAs}

\subsection{Observed Trends}
We now compare the observed Zn metallicity evolution of DLAs and sub-DLAs. 
For several reasons, the nearly undepleted element Zn is the most robust 
metallicity indicator in DLAs. Zn tracks Fe closely, to within 
$\sim  \pm 0.1$ dex, in Galactic halo and disk stars 
for [Fe/H] $\gtrsim -3$ (e.g., Cayrel et al. 2004; Chen, Nissen, \& Zhao 2004). Furthermore, Zn II has 
two, often unsaturated, absorption lines at $\lambda \lambda 2026, 2062$ 
that, for low-$z$ QSOs, are free of blending with the Lyman-$\alpha$ forest. 
\footnote{Some authors have advocated using S and Si. We prefer to use Zn 
alone to have homogeneous data 
without the complication of nucleosynthetic differences between 
different elements. Furthermore, Si can be considerably depleted in some cases 
(e.g., [Si/Zn] = -0.9 dex in the cold diffuse interstellar clouds). 
Finally, the only accessible lines of Si and S are often saturated and 
the S lines nearly 
always lie in the Lyman-$\alpha$ forest. Nevertheless, as discussed below, 
we have 
checked that using S and Si measurements in cases of Zn limits makes 
essentially no change to our results.} We do not include 
[O/H] estimates based on X-ray absorption measurements because these 
may be systematically biased with respect to [Zn/H]. 
Finally, we do not include metallicity estimates based on emission lines 
detected from absorber galaxies (available only for a few DLAs) since 
these do not necessarily probe the metallicity of the absorbing gas. 

Our samples contain 119 DLAs with $0.1 < z < 3.9$ and 30 sub-DLAs with 
$0.6 < z < 3.2$. (No Zn data exist for sub-DLAs with $z < 0.55$ because 
Zn II $\lambda \lambda 2026, 2062$ can be accessed at $z < 0.55$ only 
with space-based UV spectrographs, and the few existing UV measurements  
at $z < 0.55$ focussed on DLAs rather than sub-DLAs.) The samples are based 
on our recent data (Khare et al. 2004; 
Kulkarni et al. 2005; Meiring et al. 2006a; 2006b; P\'eroux et al. 2006b) and data from the literature. 
Recently Prochaska et al. (2006) reported 2 metal-rich sub-DLAs 
at $z \approx 1.8$, but the use of these systems for studying the  
sub-DLA metallicity evolution could be debated since they 
were discovered in a study targeted specifically toward metal-rich systems, 
i.e. those with strong Zn lines detected in the SDSS spectra. We therefore 
do not include the Prochaska et al. (2006) systems in our primary sample, 
but do discuss the effect of adding these two systems. For the same reason, 
we do not include the DLAs from Herbert-Fort et al. (2006), but discuss the effect 
of adding these DLAs later. Table 1 lists 
our primary sub-DLA sample (sample 1), together with the DLA sample. 
The DLA sample contains 68 detections and 51 limits, while the sub-DLA sample 
contains 13 detections and 17 limits. $80 \%$ of the DLA Zn limits and 
$53 \%$ of the sub-DLA Zn limits are at $z > 1.5$. 

Our analysis uses 
statistical techniques described in Kulkarni \& Fall
(2002). We divided the
119 DLAs in 6 redshift bins with 19 or 20 systems each
and the 30 sub-DLAs in
3 redshift bins with 10 systems  each. In each
bin, we calculated the
N$_{\rm H I}$-weighted mean metallicity, using the
Kaplan-Meier estimator to
account for the presence of Zn II upper limits.  These
calculations were done
using the astronomical survival analysis software
ASURV Revision 1.2 (Feigelson \& Nelson 1985; Isobe \&
Feigelson 1990). 
Fig. 1 shows the results. The circles denote DLAs, 
while the squares denote sub-DLAs (sample 1). 
The look-back times are estimated
assuming the concordance
cosmology $\Omega_{m} =0.3$, $\Omega_{\Lambda} = 0.7$,
and $H_{0} = 70$ km
s$^{-1}$ Mpc$^{-1}$.  In each bin, the data point is
plotted at the median look-back time, with horizontal bars indicating the  range
of look-back times
spanned by the systems in that bin. The vertical error
bars denote the 1
$\sigma$ uncertainties in the N$_{\rm H I}$-weighted mean metallicity, 
including sampling as well as measurement uncertainties. 

Clearly, the global mean metallicity of DLAs is
substantially sub-solar even at
low $z$ ($- 0.82  \pm 0.15$ in the bin $ 0.1 < z < 1.2$). Since the 
lowest redshift 
 bin spans a large time interval, we also 
calculated the metallicity in two halves of this bin 
with 10 DLAs each and obtained nearly identical values 
(denoted by dashed circles in Fig. 1). 
On the other hand, the two lower-$z$ sub-DLA points appear 
higher than the DLA points. The N$_{\rm H I}$-weighted mean 
metallicity in the 
sub-DLA bin at $0.6 < z < 1.4$ is $0.05  \pm
0.27$. A linear regression fit 
to the N$_{\rm H I}$-weighted mean metallicity vs. 
redshift data gives 
a slope of -0.25 $ \pm$ 0.07 for DLAs (in agreement with 
Prochaska et al. 2003a and Kulkarni et al. 2005), 
and  -0.36 $ \pm$ 0.19 for sub-DLAs. 

We next attempt to judge the dependence of our results on the 
sample definition, the 
use of Zn, and the use of $N_{\rm H I}$-weighted metallicities. We 
repeated our calculations for a number of changes in technique and assumptions. 
These results are summarized in Table 2. Concerning sample definitions, 
for example, we find that including the 
2 metal-rich sub-DLAs reported by Prochaska et al. (2006) increases  
the metallicity in the middle redshift bin by 0.17 dex. We also find that the 
DLA results change very little if the DLAs from Herbert-Fort et al. (2006) are 
included: metallicities within 0.01-0.10 dex of those shown in 
Fig. 1 for 5 of the bins, within 0.2 dex for the remaining bin at $2.6 < z < 3.1$, 
and the same linear regression slope ($-0.25  \pm 0.07$). 
If [S/H] or [Si/H] are used, where available ($z > 1.8$ and $z > 1.3$, respectively), 
in cases of Zn limits, 
and solar S/Zn or Si/Zn ratios assumed, 
the 2 higher redshift sub-DLA points in Fig. 1 change by $\le 0.05$ dex. 
(There was no 
change in the lowest redshift bin where S or Si data do not 
exist, and the relative number of Zn 
limits is anyway lower.) The resultant slope of the sub-DLA 
metallicity-redshift relation is virtually identical 
to that obtained from Zn alone, i.e., $-0.39  \pm 0.19$.  The sub-DLA 
trend suggested by Fig. 1 is, therefore, not an artifact caused by the use 
of Zn. Finally, repeating our calculations using the unweighted 
mean metallicity also gives similar results. 

It thus appears that the sub-DLA global mean metallicity 
may be higher than that of DLAs, reaching a near-solar value at
low $z$, consistent with predictions for the global mean 
metallicity in galaxies from chemical evolution models (see the curves in
Fig. 1).  Of course, it should be remembered that not every sub-DLA is  metal-rich, 
and that our result holds for the $N_{\rm H I}$-weighted mean metallicity. 
This result agrees with 
the conclusion of York et al. (2006) that the average 
[Zn/H] $\gtrsim -0.2$ 
for a composite spectrum made from 698 SDSS 
absorbers 
with average log $N_{\rm H I} \sim 20.0$. Furthermore, our result 
supports the conclusion of K06, 
based on observed correlations between [Zn/H] and $N_{\rm H  I}$, 
and a detailed consideration of various selection effects involved therein, 
that sub-DLAs are more metal-rich than DLAs. Our result also supports 
a similar suggestion by P\'eroux et al. (2003) based on [Fe/H]. 

It is important to note that the low-$z$ sub-DLAs in our sample were not 
chosen {\it a priori} to have strong Zn lines. [In fact, none of 
the 10 SDSS sub-DLAs with $z < 1.5$ in our sample showed Zn detections in 
the low-resolution SDSS spectra. Three of these 10 SDSS systems show 
detections of other relatively weak lines (Si II $\lambda$ 1808 or 
Fe II $\lambda 2249$) in the SDSS data, but interestingly, all of these 
systems have low metallicity [Zn/H] $< -0.8$.] The low-$z$ sub-DLAs 
that we observed for Zn II were chosen 
based on a known H I column density from HST UV data, moderately 
strong Fe II $\lambda 2344$ lines in the SDSS spectra, and a reasonably bright 
background quasar (usually with $V \la 19$ magnitude) with coordinates appropriate for the observing site and time. 
It is important to note that such systems are much more common than the 
``metal-strong'' systems (with strong Zn II $\lambda 2026$ or Si II 
$\lambda 1808$ lines seen in SDSS spectra) 
used in the sample of Herbert-Fort et al. (2006). 
Furthermore, the samples in Fig. 1 include data from the literature 
as well ($50 \%$ of sub-DLAs and $43 \%$ of DLAs at $0.6 < z <1.5$). 
Finally, the DLAs at $z \la 0.5$ 
from Kulkarni et al. (2005) were selected regardless of the 
Fe II strengths, and most were still found to be metal-poor. 

\subsection{Some Potential Selection Effects}

\subsubsection{Does Mg II selection bias the low-$z$ sub-DLA sample toward 
metal-rich systems?}

Most of the low-$z$ sub-DLAs and DLAs for which we have obtained Zn measurements 
were discovered in the HST UV surveys of Mg II-selected systems ($W_{2796}^{rest} \ge 0.3$ \AA; e.g., Rao, Turnshek, \& Nestor 
2006). One might, therefore, wonder whether the Mg II-selected 
systems would tend to be more metal-rich  
simply because they are selected by the presence of strong Mg II 
lines. 
If a constant rest equivalent width threshold of 
Mg II $\lambda 2796$ implied a fixed threshold of $N_{\rm Mg II}$, then one 
may think that systems with low $N_{\rm H I}$ would be included only if 
they were more metal-rich, compared to systems with higher $N_{\rm H I}$. 
However, there are several indications that this does not appear to be the 
reason for the higher metallicity we see in sub-DLAs compared to DLAs. 
First, Figs. 5-7 of York et al. 
(2006),  
based on hundreds of SDSS Mg II systems, show that strong  $W_{2796}^{rest}$ 
occur for all $E(B-V)$. This, together with the trend of 
decreasing metallicity with increasing $E(B-V)$ in the data of York et al. (2006) 
(e.g., their Table 7) implies that using systems with large  
$W_{2796}^{rest}$ does not bias one toward selecting only metal-rich 
sub-DLAs. Second, the 
Mg II $\lambda 2796$ line is almost always highly saturated in DLAs and 
sub-DLAs. The systems with larger $W_{2796}^{rest}$ have 
larger velocity spreads 
rather than larger Mg II column densities. For example, the compilation of 
Ledoux et al. 
(2006) for DLAs/sub-DLAs based on VLT UVES spectra shows that the 
velocity spread of such systems is typically about 100 km s$^{-1}$, and is 
even $\ge 200$ km s$^{-1}$ in some cases. In other words, selecting 
systems with larger $W_{2796}^{rest}$ does not necessarily translate into 
selecting systems with larger $N_{\rm Mg II}$. Finally, 
as seen in Table 1, the $N_{\rm H I}$ values of the 
sub-DLAs in our samples are relatively large: $86 \%$ of the 30 systems 
in our primary sample 
have log $N_{\rm H I} > 19.7$ and $73 \%$ have log $N_{\rm H I} > 19.9$. The 
difference in abundance needed in order to exceed a given threshold of 
$N_{\rm Mg II}$ 
is thus not substantial between our DLA and sub-DLA samples. 

Fig. 2(a) plots [Zn/H] {\it vs.} $W_{2796}^{rest}$ for DLAs and 
sub-DLAs at $z < 2$ for which both quantities are available. (We consider 
only $z < 2$ here because Mg II $\lambda 2796$ is 
not available for the $ z > 2$ systems.). Systems with large $W_{2796}^{rest}$ 
appear to show a wide range of [Zn/H], and systems with high [Zn/H] also 
show a range of $W_{2796}^{rest}$. In other words, Mg II selection does not 
appear to bias the sub-DLA sample toward only solar or supersolar systems. Note 
that the fact that the sub-DLAs in Fig. 2(a) have [Zn/H] $\ga -1$ merely states 
the observed difference between DLAs and sub-DLAs that we have mentioned all 
along in this paper. It does not necessarily indicate a metallicity bias in the 
sub-DLA sample. 

To quantify the above considerations further, we now estimate what minimum 
[Mg/H] can be covered by the low-$z$ sub-DLAs in 
our sample where the lowest $W_{2796}^{rest}$ is 0.48 \AA. We use Mg II/ H I 
to estimate [Mg/H], given that (a) most photo-ionization calculations for 
sub-DLAs predict Mg/H to differ from  
Mg II /H I by at most $0.1-0.2$ dex for the relevant range of ionization parameters 
(e.g., Dessauges-Zavadsky et al. 2003; Meiring et al. 2006b); and (b) this 
also appears to be the case for diffuse interstellar clouds
within the Milky Way (e.g. Jenkins, Savage, \& Spitzer 1986; Cartledge et al. 
2006).  If one assumes a 
linear curve of growth, then $W_{2796}^{rest} \ge 0.5$ \AA \, would correspond 
to 
log $N_{\rm Mg II} \ge 13.1$, i.e., [Mg/H] $\ge  -2.2$ for log $N_{\rm H I} = 
19.7$. Of course,  it is not correct to assume a linear curve of growth for 
such strong Mg II lines. To make more accurate estimates of the [Mg/H] threshold of 
our low-$z$ sub-DLA sample, we generated a range of synthetic Mg II $\lambda$ 2796 Voigt 
profiles with $W_{2796}^{rest}$=0.5 \AA. We considered 3 types of component velocity 
structures: 
{\bf (a)} To begin, we considered the simplest and most conservative case of  
only 1 component (extremely unlikely given 
the complex Mg II $\lambda$ 2796 line profiles observed in existing high-resolution studies of 
DLAs/sub-DLAs). For a single component with an effective Doppler $b$ parameter of 30 km 
s$^{-1}$, 
we get $W_{2796}^{rest} \ge 0.5$ \AA \, if log N$_{\rm Mg II} \ge 13.3$, i.e. 
[Mg/H] $\ge -2.0$ for log $N_{\rm H I}=19.7$. For a single component with 
effective $b$ of 20 km s$^{-1}$, we get [Mg/H] $\ge -1.7$ for 
log $N_{\rm H I} = 19.7$. 
{\bf (b)} Next, we considered a more realistic, yet simple, case of multiple narrower components. 
For 3 components with equal N$_{\rm Mg II}$ at velocities 
$v = -50, 0, 50$ km s$^{-1}$ with $b=10$ km s$^{-1}$ each, 
one would need total  log N$_{\rm Mg II} \ge 13.3$, i.e., [Mg/H] $\ge -2.0$ for log 
$N_{\rm H I}= 19.7$ to have $W_{2796}^{rest} \ge 0.5$ \AA. 
{\bf (c)} Finally, we produced profiles with 3 to 9 components with random relative 
$N_{\rm Mg II}$, $b$, and $v$ values. 
Each component was given a $b$ value ranging from 5 to 30 km s$^{-1}$ 
chosen with a random number generator. The velocities of the components 
were taken to range from -100 to 100 km s$^{-1}$, again selected with a random 
number generator. Each component was given a starting column density distributed 
randomly in the range 9 $<$ log $N_{\rm Mg II} < 11$. Then, the column 
density of each component was increased by a small constant factor in successive 
iterations until the total EW of the profile was 0.5 \AA. Simulations were repeated for 
3 to 9 components. Finally, the simulations were repeated for velocity ranges 
of -200 to 200 km s$^{-1}$ and -50 to 50 km s$^{-1}$. 
For each case, 1000 profiles were generated as described above, and histograms 
of the total Mg II column densities were examined. In these simulations, 
we obtained total median log $N_{\rm Mg II}$ in the range of 13.3 to 13.2  
(with the total $N_{\rm Mg II}$ decreasing with increasing number of components). 
In other words, one can reach [Mg/H] $\ge -2.0$ for 
$W_{2796}^{rest} \ge 0.5$ \AA \, for log $N_{\rm H I} = 19.7$. Thus, for all velocity structures considered, 
the sub-DLA sample should include systems with [Mg/H] $\ga -2$, not only 
those with [Mg/H] $> -1$. 

Of course, systems with single narrow 
components  would only be included in the $W_{2796}^{rest} \ge 0.5$ \AA \, 
sample for higher metallcities, but in general Mg II $\lambda 2796$ lines 
in DLAs/sub-DLAs are not observed  
to show such narrow velocity structures. The $N_{\rm Mg II}$ needed to 
reach higher  $W_{2796}^{rest}$ can be larger, but such systems usually also 
have larger velocity spreads than systems with lower $W_{2796}^{rest}$. 
For $W_{2796}^{rest} = 1 $ \AA, our simulations with 5 to 9 components 
spread over -100 to 100 km s$^{-1}$ or -200 to 200 km s$^{-1}$ give [Mg/H] 
in the range of -1.5 to -1.7 for log $N_{\rm H I}= 19.7$. For $W_{2796}^{rest} = 
1.5 $ \AA, the simulations with 5 to 9 components 
over the velocity range -100 to 100  km s$^{-1}$ 
give [Mg/H] of about -1.0 to -1.2 dex for log $N_{\rm H I}= 19.7$ 
or -1.3 to -1.5 dex for log $N_{\rm H I}= 20.0 $ (64 $\%$ of the sub-DLAs in 
our sample with $W_{2796}^{rest} > 1.5 $ \AA \, have log $N_{\rm H I} > 20.0$). 
Even these values are almost an order of magnitude lower than the 
metallicities actually observed for most of these sub-DLAs. In any 
case, the sub-DLAs in our sample were not specifically chosen to  have very 
large $W_{2796}^{rest}$. This is the resaon we calculate the 
[Mg/H] thresholds using the lowest $W_{2796}^{rest}$ in our sample (0.48 \AA). 
We therefore believe that while a bias against very low 
metallicity could be present in some situations (e.g., narrow single 
components or very large $W_{2796}^{rest}$), this is not likely to explain 
the difference between DLAs and sub-DLAs in Fig. 1. Indeed, as Fig. 2(a) 
shows, the higher  $W_{2796}^{rest}$ systems do not have systematically 
higher [Zn/H], and vice versa. Furthermore, as we show in section 2.3, 
our results would not be affected much even if there were a bias 
against sub-DLAs with [Zn/H] $< -1$. We note, however, that systematic 
observations of a 
much larger sample of DLAs and sub-DLAs are essential to study the 
issue of selection effects more definitively. 

\subsubsection{Observational Sensitivity for Weak Zn II Lines}

Fig. 2(b) plots log $N_{\rm Zn II}$ vs. $W_{2796}^{rest}$ for the systems from 
Fig. 2(a). The DLAs show a narrower range of 
$N_{\rm Zn II}$ than the sub-DLAs. In fact, the DLAs and sub-DLAs with 
comparable $W_{2796}^{rest}$ do not 
necessarily have similar $N_{\rm Zn II}$. There are many $N_{\rm Zn II}$ 
detections or upper limits for sub-DLAs that are actually {\it below} the DLA 
points at comparable $W_{2796}^{rest}$. Clearly, the sub-DLA measurements have 
adequate sensitivity to see low Zn II column densities.  The 
{\it metallicities} of some of the sub-DLA systems, derived after the fact by dividing 
the $N_{\rm Zn II}$ by $N_{\rm H I}$, are high (see Fig. 2a). The $N_{\rm H I}$ values play no direct role in the detection of Zn II lines. 
 Of course much stronger Zn II lines could have been detected as well, and would in fact, have to be there if there were a 
significant number of solar or supersolar metallicity DLAs. However, such strong Zn II lines for DLAs are not seen. On the other hand, there 
are a small fraction of sub-DLAs that have $N_{\rm Zn II}$ larger than 
those for many DLAs, and these are of course highly metal-rich systems. 
Thus, the larger fraction of metal-rich sub-DLAs does not appear to be 
a result of sampling similar $N_{\rm Zn II}$ values at lower $N_{\rm H I}$ 
values compared to the DLAs. 

\subsection{The Role of Metal-rich and Metal-poor Systems}

It is interesting to compare the mean metallicities of the subsets of the DLA 
and sub-DLA samples after removing the systems with [Zn/H] $< -1$, since there 
cannot be a selection 
bias against the remaining systems. \footnote{Of course, given the discussion above, 
the sub-DLA samples should be sensitive well below [Zn/H]$ = -1$ dex.} 
These subsamples consist 
of 50 DLAs and 27 sub-DLAs. Dividing these 
subsamples into 3 bins each, the $N_{\rm H I}$-weighted mean metallicity is 
$-0.66 \pm 0.15$ in the lowest redshift DLA bin, while it is $0.08 \pm 0.28$ 
in the lowest redshift sub-DLA bin. While these numbers are certainly not 
representative of the overall observed DLA and sub-DLA populations (since by 
design they exclude the metal-poor systems), the DLA and sub-DLA 
means still seem to differ at $2-3 \, \sigma$ level. {\it Thus, the difference between 
DLA and sub-DLA metallicities at low $z$ would not be affected much even if the sub-DLA 
samples were biased against systems with [Zn/H] $< -1$.}

In principle, if there were a large 
number of metal-poor sub-DLAs with metallicity $\la 1/100$ solar that 
could be missing from the Mg II-selected samples, then the ``true'' sub-DLA mean 
metallicity could be lower. To assess this possibility, we repeated the 
survival analysis calculations after adding a hypothetical metal-poor 
sub-DLA population with a typical log $N_{\rm H I} = 20$ in the lowest 
redshift bin (where the difference between DLA and sub-DLA mean metallicities is largest).
We conclude that one would need to add 64 sub-DLAs with [Zn/H] = -2.0  
(or 75 sub-DLAs with [Zn/H] = -1.5) in order to bring the $N_{\rm H I}$-weighted 
mean metallicity of the sub-DLAs down to the DLA level (-0.82 dex) 
in the lowest redshift bin. Such a large population of 
metal-poor sub-DLAs cannot be ruled out, but seems quite unlikely, given that 
there are currently only 10 
sub-DLAs with Zn measurements in this redshift bin (of which 6 are solar or 
supersolar). Turning the argument around, if 
the fraction of metal-poor sub-DLAs were the same as the fraction of DLAs with 
[Zn/H] $< -1$, e.g., if there were 15 hypothetical sub-DLAs 
with a typical [Zn/H] = -1.5 in addition to  
to the 10 observed sub-DLAs in the lowest redshift bin, then the  $N_{\rm H I}$-weighted 
mean metallicity of the combined sample of 25 sub-DLAs would be -0.35 dex, 
not -0.82 dex as for DLAs. 
In other words, it would take a bimodal metallicity distribution 
very different from that for known DLAs to force the sub-DLA mean to equal the observed 
DLA mean. In fact, since the low-$z$ DLAs are also discovered in the Mg II-selected 
HST surveys, in principle there could be a population of low-$z$ DLAs 
with [Mg/H] $< -2.6$ undetected in current DLA surveys that could 
drive the DLA mean metallicity to even lower values. Thus the DLA and sub-DLA means could 
still remain different. 

{\it It appears that the primary reason for the difference between the observed  
DLAs and sub-DLAs in Fig. 1 is not a bias against metal-poor sub-DLAs, 
but the dearth of metal-rich DLAs, for which there should be no difficulty in 
detecting the metal lines.} Fig. 1 shows that metal-rich systems can be found 
among the sub-DLAs, whereas they seem to be quite rare among the DLAs, 
despite much more extensive observations for the DLAs.  
Larger samples are needed, at any rate, to refine the 
differences between metal-rich sub DLAs and metal-poor sub-DLAs and 
DLAs, and those larger samples can be searched for more subtle or 
different selection effects than those we can study here.

\subsection{Other Considerations}

We note that owing to the relatively small size of the sub-DLA 
sample, the  mean metallicities for this sample are sensitive to 
the particular details of the metal-rich sub-DLAs discovered so far. The maximum effect on the 
mean metallicity in the lowest-$z$ bin comes from the sub-DLA with the 
highest $N_{\rm Zn II}$ in our sample, i.e., the system at $z = 0.716$ toward SDSSJ1323-0021 
(P\'eroux et al. 2006a). We have adopted log $N_{\rm H I} 
= 20.21^{+0.21}_{-0.18}$ for this system since this gives the best 
fit to the Ly-$\alpha$ absorption profile 
(Khare et al. 2004; P\'eroux et al. 2006a). Rao et al. (2006)  
reported log $N_{\rm H I} = 20.54 \pm 0.15$ for this system, while Prochaska 
et al. (2006) reported log $N_{\rm H I} = 20.3$. The value of Prochaska et al. 
is only stated to one decimal, without error bars, but is much closer to 
the value adopted here. In any case, if one were to take this system 
out of the sub-DLA sample, and put it instead in the DLA sample 
(adopting the Rao et al. 2006 value for its $N_{\rm H I}$), then the sub-DLA mean 
metallicity in the lowest redshift bin would become $-0.26 \pm 0.16$. 
The corresponding value in the DLA sample would change only a little to 
$-0.73 \pm 0.15$. (The DLA sample is larger and is less affected by an 
individual system.) Thus, the mean metallicity at low redshift would 
be still higher for the sub-DLA sample than for the DLA sample 
at the $3 \sigma$ level, but the difference would be smaller.  

Overall, we conclude that the higher mean sub-DLA metallicity does not appear 
to be an artifact of sample selection (on the basis of either Mg II or 
other metal lines), the use of Zn, or the use of $N_{\rm H I}$-weighting, and 
does not arise from insufficient  
$N_{\rm Zn II}$ sensitivity. However, it is necessary to verify our 
results with (a) much larger sub-DLA and DLA Zn samples, and (b) more accurate 
$N_{\rm H I}$ measurements for DLAs as well as sub-DLAs with higher 
resolution and higher S/N observations of the Ly-$\alpha$ profiles.

\section{ Contribution of Sub-DLAs to the Cosmic Metal Budget} 

Our knowledge of the cosmic metal budget is highly incomplete. 
Earlier estimates including contributions of high-$z$ DLAs, Lyman-break galaxies, 
and Lyman-$\alpha$ forest suggested only 
$\sim 10 \%$ of the amount of metals expected from the cosmic 
star formation history (e.g., Pettini 2004).  
Recent estimates also including sub-mm galaxies and star-forming galaxies 
at $z \sim 2$ yield $\sim 
40-50 \%$ of the expected  
comoving metal density ($\approx 4 \times 10^{6}$ M$_{\odot}$ Mpc$^{-3}$; 
Ferrara, Scannapieco \& Bergeron 2005; Bouch\'e, Lehnert 
\& P\'eroux 2005, 2006a; Pettini 2006). 
It has been suggested that dwarf galaxies contaminate the intergalactic 
medium with hot ionized gas (e.g., Tumlinson \& Fang 2005), 
but the extent of this effect is not clear. 
Given the available observations, about half of 
the predicted quantity of metals appears to be unaccounted for. 

In this context, it is interesting to quantify the
contribution of the
possibly metal-rich sub-DLAs to the metal budget. 
At low redshift it is difficult to estimate the
sub-DLA contribution 
with the available data. The statistics for sub-DLAs
and hence the comoving density of H I gas  $\Omega_{\rm HI}$ in 
sub-DLAs are not yet known at 
$z < 1.7$. However, one could assume that the
relative H I contributions
of DLAs and sub-DLAs at low $z$ are similar to
those at high $z$. At 
$z\sim$2.5, the co-moving density of
H I gas in DLAs and sub-DLAs is measured to be $\Omega_{\rm
H\;I}$= 0.85 $\times $
10$^{-3}$ and 0.18 $\times$
10$^{-3}$, respectively (P\'eroux et al. 2005), i.e., a ratio 
 $\Omega_{\rm
H\;I}^{\rm sub-DLA} / \Omega_{\rm
H\;I}^{\rm DLA} = 0.21$. Therefore, the co-moving mass
density of metals in sub-DLAs, in units of
$\Omega(Z_{\odot})$ ($=\Omega_{\rm
baryons} \times Z_{\odot}=5.5 \times 10^{-4}$;
$Z_{\odot}$=0.0126 by mass),
is: $\Omega_Z^{sub-DLA}=f\times 10^{0.0} \times
Z_{\odot} \times  \Omega_{\rm
H\;I}^{\rm sub-DLA}/ \Omega(Z_{\odot})$=$f\times\; 22.9  \times \Omega_{\rm
H\;I}^{\rm sub-DLA}$ where $f$ ($>$1)
accounts for the ionized fraction of the gas. For
sub-DLAs near the high $N_{\rm H I}$ end, the ionized fraction is not
important, but it may be
important for sub-DLAs with low $N_{\rm H I}$.
The exact value of $f$
will depend on detailed photoionization calculations
as well as the $N_{\rm H I}$ distribution of the sub-DLAs, but is 
likely to be in the range 1-10. Thus, 
if the sub-DLA metallicity exceeds that 
of DLAs by a factor of about 7 at low $z$ (Fig. 1), and the
relative H I contributions of DLAs and sub-DLAs at low $z$ are similar to
those at high $z$, then the metal contribution of sub-DLAs 
will exceed that of DLAs by a factor of $f\times$ 1.5 
[since $\Omega_Z^{DLA}=2.6 \times
10^{-3}$, in units of $\Omega(Z_{\odot})$]. 
This factor may be even higher if the relative H~I contribution
of sub-DLAs at low $z$ is higher than that at high $z$ 
(e.g., the factor will be $f\times$ 4.5 if 
$\Omega_{\rm HI}$ values of Prochaska et al. 2005 
are used). 
These
expectations will need to be confirmed with more data 
on sub-DLA $\Omega_{\rm
H\;I}$ and metallicity.

At high redshifts, the mean global metallicity of
the sub-DLAs is $[\langle Zn/H \rangle] \approx -0.7$ (Fig. 1).
Therefore, the co-moving mass
density of metals in sub-DLAs, in units of
$\Omega(Z_{\odot})$, 
is: $\Omega_Z^{sub-DLA}=f\times 10^{-0.7} \times
Z_{\odot} \times 0.18 \times
10^{-3} / \Omega(Z_{\odot})=f\times\;8.2 \times
10^{-4}$.  Thus, at high redshifts, sub-DLAs
contribute at least 32\%
of what DLAs contribute to the metal census. 
The sub-DLA contribution could be higher  depending on the
value of $f$. 
Prochaska et al. (2005) based on their
analysis of the Sloan Digital Sky Survey (SDSS) Data Release 3 (DR3), find a
value of $\Omega_{\rm H\;I}$ for systems with $17.3 \le \rm{log} \, N_{\rm
H I} \le 20.3$ to be 0.57
times $\Omega_{\rm H\;I}$
in DLAs, a large fraction coming from the sub-DLAs.
Thus, the sub-DLA
contribution to the metal budget may be higher than
that estimated above. We
note, however, that the conclusions of Prochaska et al. (2005) will need
to be confirmed at high resolution, which is essential to 
obtain reliable N$_{\rm H\;I}$ for sub-DLAs (P\'eroux et
al. 2005). Taken together, the suggestions of high metalicities at $z \sim 2$ 
for a few sub-DLAs and our finding of an even larger effect at 
low redshifts are consistent with sub-DLAs being more significant contributors 
to the total metal abundance of gas in the Universe than was previously realized. 

\section{Summary and Discussion}

While the sub-DLA samples are still small and not all 
the observed sub-DLAs are metal-rich, the fraction of observed  
metal-rich systems is larger for sub-DLAs than for DLAs. Our calculations 
suggest that the mean 
metallicity of the observed sub-DLAs is higher than that of the observed 
DLAs, reaching a 
near-solar level as expected in cosmic chemical evolution 
models. At $z < 1$, we find that the contribution of the observed sub-DLAs to the 
total metal budget may be several times higher than that of the observed 
DLAs. At higher $z$, the observed sub-DLAs may contribute $\ge 32 \%$ 
of the DLA contribution to the total metal budget. Furthermore, the 
results of Prochaska et al. (2005) on the relative number of sub-DLAs and 
DLAs may triple the contribution of sub-DLAs compared to the results 
based on P\'eroux et al. (2005). In general, 
our results support and extend the suggestions by P\'eroux et al. (2006a), York 
et al. (2006), and Prochaska et al. (2006) that sub-DLAs may contribute significantly to the overall global metallicity.

One may wonder whether ionization effects are likely to be responsible for 
the difference we see between the global metallicities for the DLA and 
sub-DLA samples. We believe that 
this is not a significant effect for our samples. First, the 
$N_{\rm H I}$ of the sub-DLAs in our sample are relatively high ($86 \%$ 
systems with log $N_{\rm H I} > 19.7$ and $73 \%$ with log 
$N_{\rm H I} > 19.9$ in our primary sample.) Second, photoionization 
calculations presented in high-resolution studies 
of sub-DLAs (Dessauges-Zavadsky et al. 2003; Meiring et al. 2006b) show 
that while sub-DLAs have some ionized gas, the ionization corrections 
in abundance estimates are not significant, and would increase [Zn/H] if applied. 
Vladilo et al. (2001), Dessauges-Zavadsky et al. (2003), York et al. (2006), 
and Meiring et al. (2006b) all suggest that there is no trend in $N_{\rm Al \, III}$/$N_{\rm Al \, II}$ vs. $N_{\rm H I}$ 
in sub-DLAs and that the ratio is smaller than 1. 
  Therefore, using $N_{\rm Zn II}/ N_{\rm H I}$ 
  in sub-DLAs for a measure of $N_{\rm Zn}/N_{\rm H}$ does not introduce a 
  systematic bias. 

It is important to establish whether the observed DLA and sub-DLA 
samples reflect their true properties or differential 
dust selection effects. If dust obscuration is significant and 
depends primarily on the metal column density (e.g., Vladilo et al. 2006), 
then metal-rich DLAs 
may be more affected by dust selection than metal-rich sub-DLAs 
(Vladilo \& P\'eroux 2005).    
Such an effect could make DLAs appear less metal-rich than sub-DLAs; 
but, this alone would not explain the large difference 
between DLAs and sub-DLAs seen in Fig. 1, because so far there is no 
evidence for the necessary large number of high extinction DLAs (K06). 
K06 use this result and the observed mass-metalicity relationship for galaxies 
(e.g., Tremonti et al. 2004; Savaglio et al. 2005; Erb et al. 2006)
  to conclude that sub-DLAs may arise in massive galaxies and DLAs  
  in less massive galaxies, consistent with the relatively stronger  
  evolution of metals in sub-DLAs compared  
  to DLAs that we find here. That conclusion is also 
  consistent with trends in gas dispersion with metalicity of quasar absorbers   
  (Ledoux et al 2006, Meiring et al. 2006b). [But, see an opposing opinion on 
  this last point by Bouch\'e et al. (2006b), using a less direct argument.] 

  The relatively smaller sample size for sub-DLAs  compared to DLAs and  
  the conflicting results on gas velocity dispersion just referred to  
  emphasize the need for much larger samples of sub-DLAs and better  
  velocity dispersion measurements for the absorbers at a range of redshifts,  
  to confirm the results reached here. Large telescopes are essential to obtain adequate  
  samples of excellent abundance and velocity dispersion measurements. 
  The sub-DLA sample needs to be made at least as  
  large as the DLA sample. It will also be important to obtain $N_{\rm H I}$ 
  and [Zn/H] for absorbers toward UV-faint quasars, to confirm  
  that there are not a large number of highly reddened quasars with 
  foreground DLAs. 
  More extensive observations with the HST Cosmic Origins 
Spectrograph to measure H I at $z < 1.5$ and Zn II  at 
$z < 0.55$, and with 
ground-based spectrographs to measure Zn II at $0.55 < z < 1.5$ will be very 
important to confirm the findings presented here . 

\acknowledgments

VPK and JDM acknowledge support from NSF  
grants AST-0206197 and AST-0607739. PK acknowledges support from the Department of 
Science and Technology, Government of India (SP/S2/HEP-07/03). We thank 
Max Pettini, Nicolas Bouch\'e, and an anonymous referee 
for comments that helped to improve this paper.

\clearpage
\begin{table}
\begin{center}
\caption{Zn Measurements in Sub-DLAs and DLAs\label{tbl-1}}
\begin{tabular}{lllllllll}
\tableline
\tableline
Quasar	& Ref.$^{a}$ & $z_{abs}$ &[Zn/H]	& $N_{\rm H I}/10^{20}$& $N_{\rm Zn II}/10^{12}$&[X/H]$^{b}$ & W$_{2796}$ & W$_{2600}$\\
&&&&(cm$^{-2}$)&(cm$^{-2}$)&&({\AA})&({\AA})\\
\tableline
&&&&&&&&\\
\bf{Sub-DLAs:}&&&&&&&&\\	
&&&&&&&&\\	
0058+019&	1&	0.6125&	0.13&	$1.20 \pm 0.42$& $6.46 \pm 1.61	$& ...&1.63&1.27\\
1028-0100&	2&	0.6321&	-0.04&	$0.79 \pm 0.26$& $2.88 \pm 1.31$&...&1.58&1.14\\	
1028-0100&	2&	0.7088&	$< 0.06$&	$1.02 \pm 0.33$& $< 4.71$&...&1.21&0.89\\	
1323-0021&	3&	0.7160&	0.62&	$1.62 \pm 0.78$& $26.80 \pm 2.77$&...&2.23&1.45\\
0134+0051&      4&	0.8420&	$< -0.36$&	$0.85 \pm 0.23$& $< 1.48$&...&1.17&0.86\\
\tableline
\end{tabular}
\tablecomments{Table 1 is presented in its entirety in the electronic edition of 
the {\it Astrophysical Journal}. A portion is shown here for guidance regarding 
its form and content.}
\tablenotetext{a}{References: 1. Pettini et al. (2000); 2. Khare et al. (2004); 3. 
P\'eroux et al. (2006a); 4. P\'eroux et al. (2006b).}
\tablenotetext{b}{ Abundances of S or Si if available, are listed for 
sub-DLAs in cases with Zn upper limits.}
\end{center}
\end{table}

\begin{table}
\begin{center}
\caption{Binned Sub-DLA Mean Metallicities for Various Samples\label{tbl-3}}
\begin{tabular}{lllll}
\tableline\tableline
Sample, & Elements & $N_{\rm H I}$- & Redshift Ranges of 3 bins$^{a}$ & Mean Metallicity in the 3 bins  
 \\
No. of & Used &weig-& \\
Systems&&hted?&&\\
\tableline
&&&&\\
1$^{b}$, 30 & Zn & Y & 0.61-1.14,1.15-1.87,1.89-3.17&
$\,\,\,\,\, 0.05  \pm 0.27$,$-0.60  \pm 0.25$,$-0.67  \pm 0.16$\\
2$^{c}$, 32 & Zn & Y & 0.61-1.15,1.22-1.87,1.89-3.17 &
$\,\,\,\,\, 0.02  \pm 0.25$,$-0.43  \pm 0.27$,$-0.78  \pm 0.21$\\
3$^{d}$, 30 & Zn,S,Si & Y & 0.61-1.14,1.15-1.87,1.89-3.17&
$\,\,\,\,\, 0.05  \pm 0.27$,$-0.55  \pm 0.23$,$-0.70  \pm 0.18$\\
1$^{e}$, 30 & Zn & N & 0.61-1.14,1.15-1.87,1.89-3.17 &
$\,\,\,\,\, 0.16  \pm 0.19$,$-0.67  \pm 0.20$,$-0.63  \pm 0.15$\\
\tableline
\end{tabular}
\tablenotetext{a}{In each case, the sample was divided into 3 redshift bins with 
roughly equal number of systems.}
\tablenotetext{b}{Primary Sample, Zn only}
\tablenotetext{c}{Sample including Prochaska et al. (2006)}
\tablenotetext{d}{Primary Sample, S or Si for Zn limits}
\tablenotetext{e}{Primary Sample with unweighted means}
\end{center}
\end{table}

\clearpage

\begin{figure} 
\includegraphics[width=3.15in,height=2.7in]{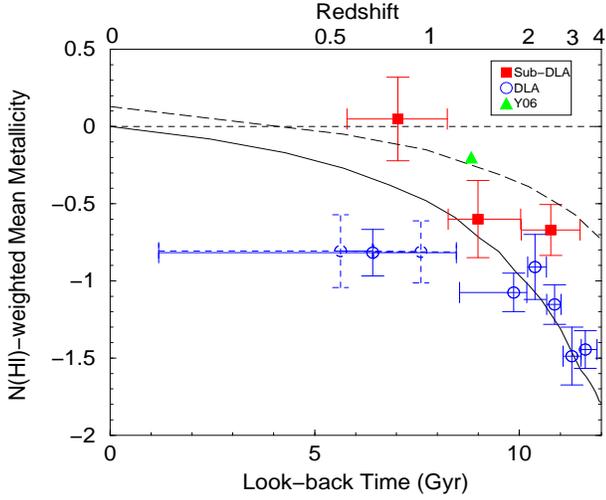}
\caption{Logarithmic $N({\rm H \, I})$-weighted mean Zn
metallicity plotted vs. look-back time 
for DLAs and sub-DLAs. Dashed circles refer to the 
lowest time bin split into 2 bins with 10 DLAs each. 
The triangle denotes the formal lower limit to the average [Zn/H] 
for a composite spectrum from 698 
absorbers with average log $N({\rm H \, I}) \sim 20$ (sample 24) 
from York et al. (2006). 
The solid and long-dashed curves show,
respectively, the mean metallicity in
the chemical evolution models of Pei et al. 
(1999) and Somerville et al. 
(2001). } 
\end{figure} 

\begin{figure} 
\plottwo{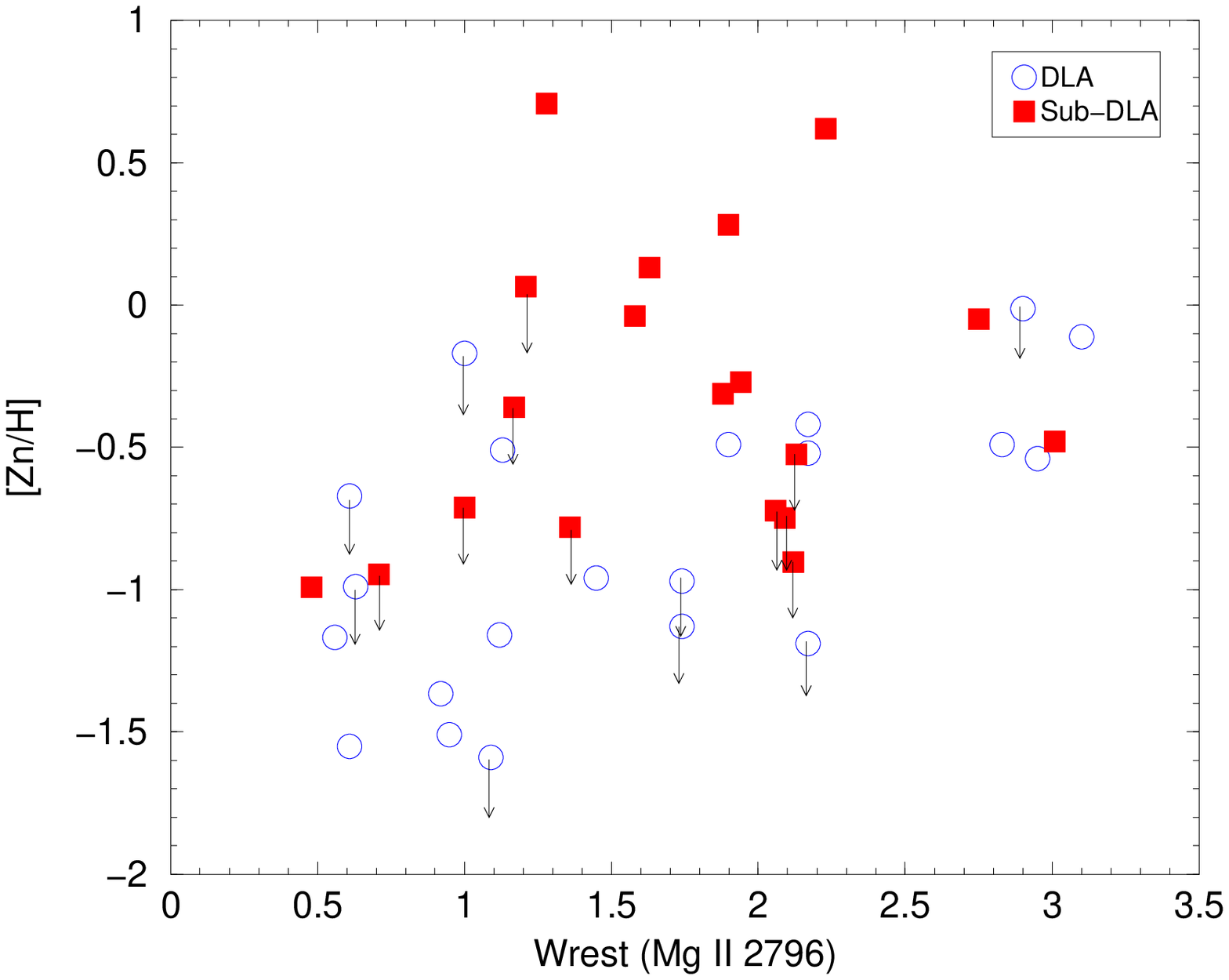}{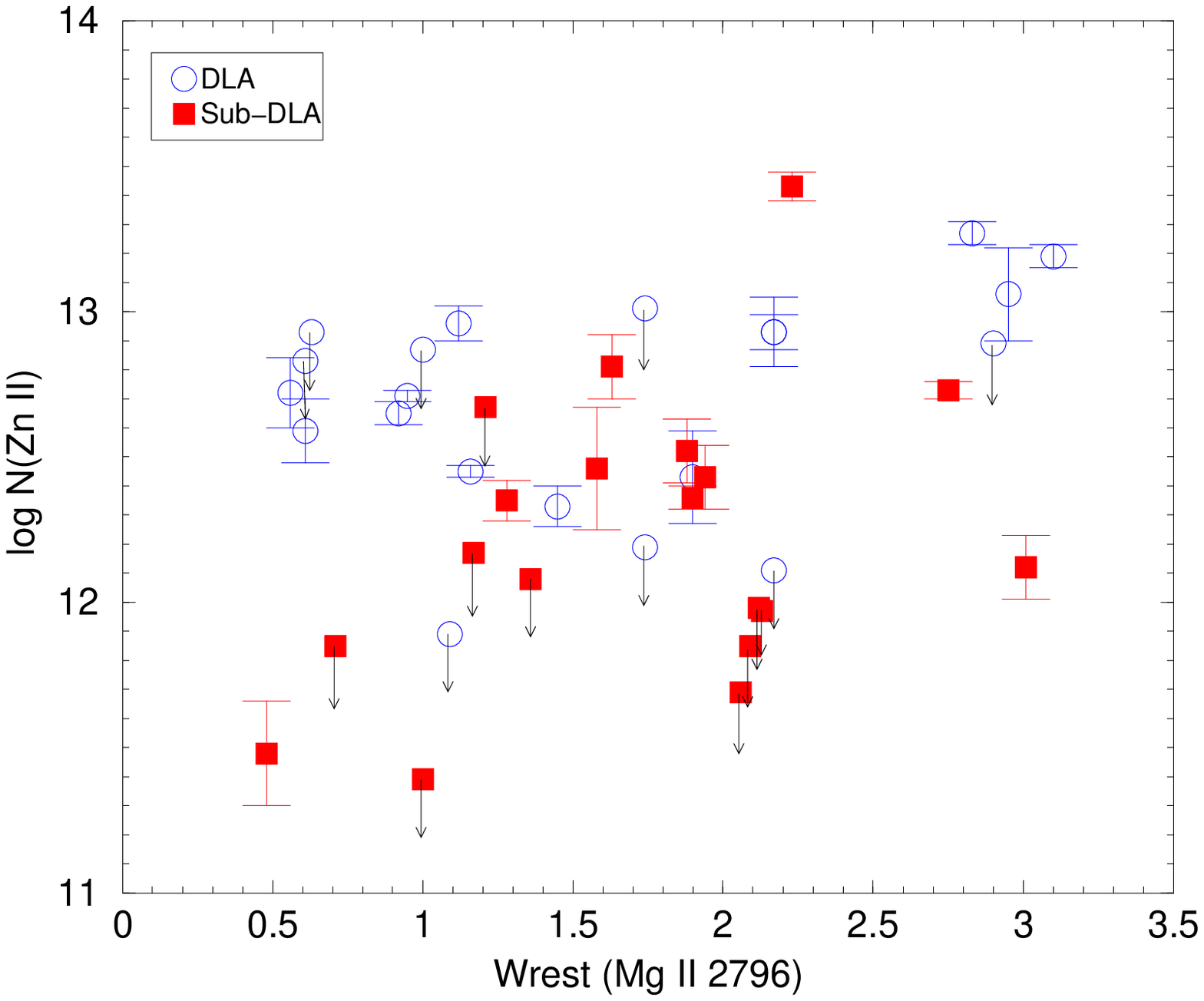}
\caption{Left (a)  [Zn/H] vs. rest equivalent width of Mg II $\lambda 2796$ 
for sub-DLAs and DLAs at $z < 2$ from our sample for which both quantities 
are available. Right (b): log $N_{\rm Zn II}$ vs.  $W_{2796}^{\rm rest}$ 
for the same set of sub-DLAs and DLAs. (See text for further details.)} 
\end{figure} 

\end{document}